\documentclass[preprint,prX]{revtex4}
\usepackage{epsfig}
\usepackage{graphicx}
\usepackage{bm,wasysym}

\begin{document}

\title{A versatile and reliably re-usable ultrahigh vacuum viewport}

\date{\today}
\author{K. J. Weatherill} \email{k.j.weatherill@durham.ac.uk}
\author{J. D. Pritchard}
\author{C. S. Adams}
\affiliation{Department of Physics, Durham University, DH1 3LE, UK.}
\author{P. F. Griffin}
\author{U. Dammalapati}
\author{E. Riis}
\affiliation{Department of Physics, University of Strathclyde, G4 0NG, UK}

\begin{abstract}
We present a viewport for use in Ultra-high vacuum (UHV) based upon
the preflattened solder seal design presented in earlier work, Cox {\em et al.} Rev. Sci. Inst. \textbf{74}, 3185 (2003). The design features significant modifications to
improve long term performance. The windows have been leak tested to less than 10$^{-10}$ atm cm$^3$/s . From atom number measurements in an optical dipole trap loaded from a vapor cell magneto-optical trap (MOT) inside a vacuum chamber accommodating these viewports, we measure a trap lifetime of 9.5~s suggesting a pressure of around $10^{-10}$~Torr limited by background Rubidium vapor pressure. We also present a simplified design where the UHV seal is made directly to a vacuum pipe.
\end{abstract}
\maketitle

A wide variety of experiments require high quality UHV viewports.
Often, those available commercially are expensive and
have limited optical quality. The sealing process for commercial viewport often distorts the edges of the substrate and anti-reflection coatings do not reach the edges of the optic due to shadowing effects thus reducing the viewing area. Consequently there has been abundant
interest in the development of alternatives with better optical quality and lower cost \cite{manuccia1981,muller1988,nobel1994}.
In previous work \cite{cox2003} we described a re-usable viewport
design based upon a solder seal and bakeable to 240$^\circ$C and
highlighted the differences and advantages of this method over other
schemes. Subsequently, this design was successfully used on cold atom experiments \cite{griff05,griff06}.
However, the motivation for further improvements arose after it was
found the solder seal developed small leaks at the $10^{-9}$~atm~cm$^3$s$^{-1}$ level
after repeated removal and reattachment to the vacuum system. This
failure was attributed to deformation of the Conflat flange on which
the solder and copper gasket seals are made due to the high stresses necessary to deform the copper gasket.

The first modified design presented here incorporates an additional flange to
decouple the solder seal from the conflat flange which is bolted
to the vacuum system. Figure~1 shows a plan view and two
cross-sectional views of the modified viewport. The viewport is
constructed by mounting an optical element on a double flange, in
this case made from a ``cut down" 70CF half--nipple with an
additional flange on which the optic is mounted manufactured in house and welded in place. This additional flange has outer diameter (OD) of 76~mm and inner diameter (ID) of 38~mm to match the ID of the half-nipple tube. There are also 12 bolt holes; 6 \diameter 11~mm clear holes are for the M6 cap screws used to tighten the conflat seal which couples the viewport to the vacuum chamber, the 6 M6 tapped holes are for attaching the clamping flange.
In principle this structure could be machined from a single billet,
thus avoiding the need for welding. The clamping flange is also manufactured in house and has OD 76~mm. It has a recess in which the substrate can fit and 12 bolt holes; 6 clear and 6 through as in figure~1. This design allows the copper
gasket seal of the Conflat flange to be made without applying stress
to the solder seal surfaces. The materials and processes of the sealing procedure are the same as
reported in earlier work. However, such is the care required to successfully seal these viewports, a step by step procedure follows. Note that all equipment which comes into contact with surfaces to be placed under vacuum should be cleaned appropriately beforehand.

The ends of the solder wire (\diameter 0.7~mm Indalloy 165 \cite{indalloy}) were carefully joined to form a loop of appropriate diameter and then
the solder ring was squashed to a thickness of approximately 0.3~mm between two flat polished surfaces. Excess material from the join area was removed with a scalpel to form a neat circle on the inside and outside edges of the ring and then the ring was squashed a second time to remove any roughness caused by the scalpel cut. At this point the join in the solder ring should appear seamless. A solder ring was then placed carefully onto the flat surface on which the vacuum seal is to be made and then the viewport substrate placed on top. A second solder ring was then placed on top of the substrate to cushion the clamping flange, which is placed on top. Two conical disc springs \cite{belleville} were placed base-to-base onto each M6 (1~mm pitch) securing bolt and the bolts were tightened to a torque of 3 Nm. The purpose of the  disc springs is to maintain the force on the clamping flange when the solder softens during baking. After cooling the bolts were re-tightened to a torque of 3~Nm after which no further tightening was required, even after further baking cycles.

\begin{figure}[h!!]
\centering
\includegraphics[width=6.7cm,angle=0]{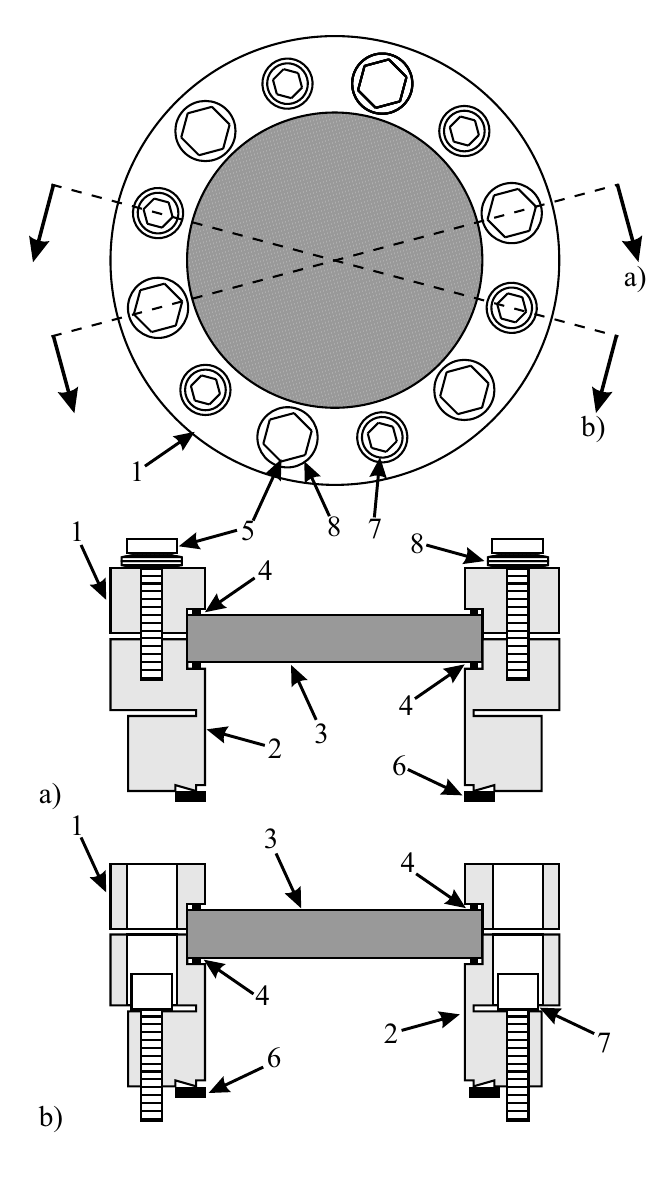}
\caption{(Top) The viewport in plan view. (Middle and lower parts)
Different cross-sectional views through the viewport, indicated by
(a) and (b). The viewport consists of: (1) clamping flange, (2)
double flange, (3) optical element, (4) preflattened Indalloy 165
solder seal, (5) hexagonal head securing bolt, (6) OFHC copper
gasket, (7) socket-head bolt and (8) conical disk springs.}
\end{figure}

During previous attempts at viewport manufacture we found that the
solder seal was less dependable when using anti-reflection (AR)
coating on substrates because the rough edge of the coating could prevent the vacuum seal being successfully made. For this reason, we used some substrates with a coating
free edge, on which to make the vacuum seal. A further refinement is the use of different bolt heads
for the viewport and knife-edge seals to eliminate accidental
over-tightening of the solder seals when the conflat seal is made. The viewports were leak tested using a Stanford Instruments RGA100 residual gas analyzer with an electron multiplier. No leaks were detected above the noise floor of the detector, $<10^{-10}$ atm cm$^3$/s. This viewport design has been repeatedly tested using BK7 and ZnSe optical elements. The viewports have been
re-attached and baked six times up to 250$^\circ$C without adversely affecting
the UHV viewport seal.

The design outlined above was implemented on viewports attached
to a vacuum chamber for use in a recent cold atom experiments \cite{weat08} and in addition used for an optical dipole trap experiment.
Figure~2 shows the number of Rb atoms trapped in
a CO$_2$ laser optical trap loaded directly from a vapor cell MOT within the chamber as a function of time. Such a far-detuned optical dipole trap is conservative \cite{grimreview} and the measured 1/${\rm e}$
lifetime of 9.5~s is limited by the background pressure in the UHV
chamber. The lifetime suggests a pressure of around 10$^{-10}$~Torr and is limited by the gas load from the Rb dispensers and pumping speed from the 40 l/s Varian diode ion pump, not the leak rate of the viewports.
\begin{figure}[hbt]
\centering
\includegraphics[width=8.0cm,angle=0]{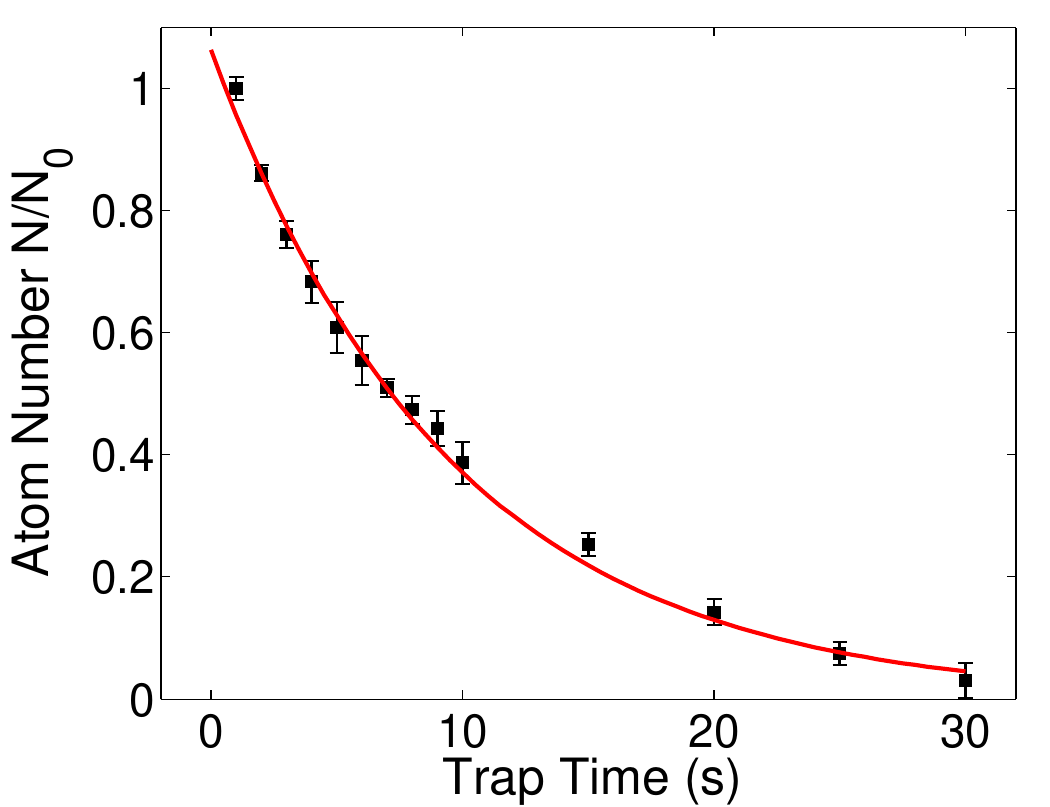}
\caption{Measurement of the number of Rb atoms in a CO$_2$ laser optical dipole trap for increasing hold time. A 1/{\rm e} lifetime of 9.5 $\pm$ 0.5~s is deduced from an exponential fit to the data.}
\end{figure}

The specialized requirement in the design presented here is in the manufacturing process. Non-standard vacuum components are used including, in the present design, the mounting flange welded onto a standard, albeit shortened, conflat half nipple. The design is also bulkier than many viewports which presents a disadvantage for optical experiments requiring high numerical aperture (NA). To address these issues we have also investigated two simplified designs. First of all we note that the key to the robustness of the design presented so far in comparison with our previous design \cite{cox2003} is the use of a length of tube between the conflat flange, that attaches to the rest of the vacuum system, and the optic to metal seal. As this seal is typically only about 2 mm wide we implemented the design shown in Figure~3, where we have discarded the mounting flange and sealed directly on the end of the tube. For a standard 70CF half nipple the tube has a wall thickness of 1.5 mm. Prior to sealing, the end of the pipe was lapped to ensure it was completely flat. The sealing process follows the same principles as outlined above. In this case the clamping flange was tightened onto the conflat flange with 12 M4 (0.7~mm pitch) bolts with spring washers. A torque of about 1.2~Nm was sufficient to ensure a good seal. Figure 3 shows the design of this window including the specially designed clamping flange. This viewport design has been leak tested to $<10^{-10}$ atm cm$^3$/s  and we have successfully operated a vacuum system with eight windows of this design at a pressure below $10^{-8}$ Torr for an extended period of time.

In order to increase the robustness of this design we have investigated the use of an ultra-high vacuum compatible epoxy to improve the glass-to-metal seal. The shaped solder ring between the steel tube and the window is maintained in order to allow for the differential expansion of the stainless steel tube and the window material. However, a thin layer of epoxy is applied on either side of the ring to ensure a strong bonding of the solder ring to both the steel tube and the window. To further increase the adhesion between the solder ring and the steel, the tube end was not lapped flat but left with a spiral shaped machine mark on the end face.
The epoxy used is Epotek 353ND, which has an exceptionally low outgassing rate \cite{epotech} and an operating temperature of up to 250$^\circ$C. For a bake to 200$^\circ$C the differential expansion between a 40 mm diameter steel tube and a quartz window is ~120~$\mu$m. Using the shear modulus for lead (97.5\% of Indalloy 165 is lead) the corresponding shear force is found to be approximately an order of magnitude less than the shear strength of the epoxy. Thus, we are confident that the solder ring deforms to account for the thermal expansion while maintaining the integrity of the actual vacuum seals between the solder ring and the steel tube and window respectively.
This technique provides an extremely reliable seal because as the viewport is baked, the epoxy becomes supple and fills any imperfections in the sealing surfaces or AR coating.
\begin{figure}[hbt]
\centering
\includegraphics[width=6.7cm,angle=0]{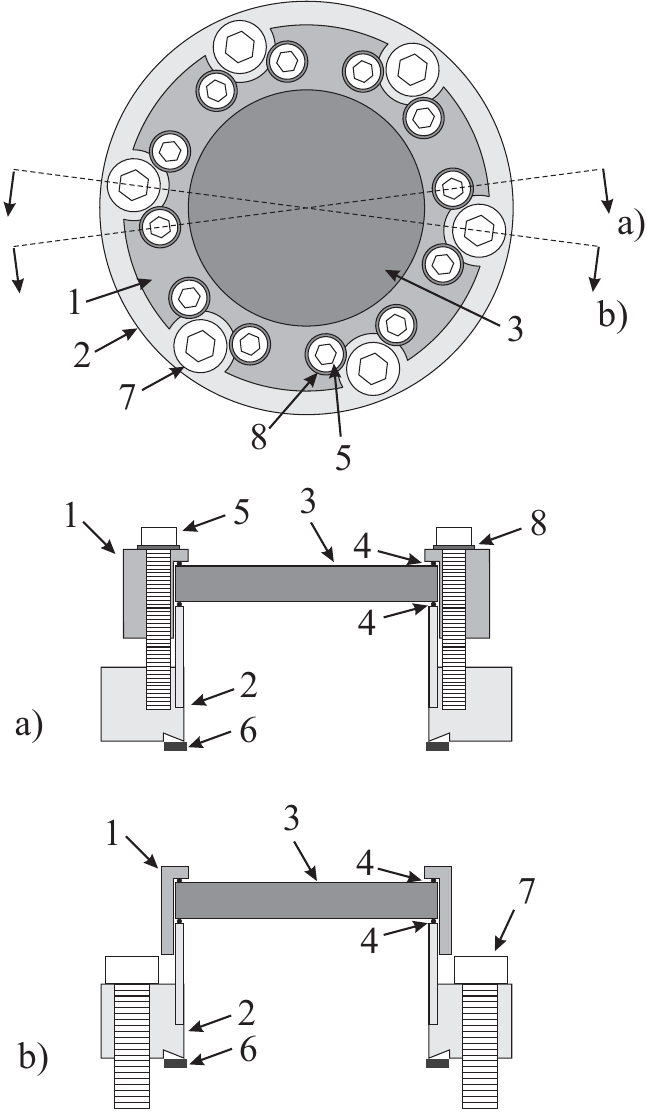}
\caption{(Top) The viewport in plan view. Below are shown different cross-sectional views through the viewport indicated by a), b) and c). The viewport consists of: (1) brass clamping flange, (2) standard but shortened stainless steel conflat half nipple, (3) optical element, (4) pre-flattened Indalloy 165 solder seal, (5) M4 socket-head bolt, (6) OFHC copper gasket, (7) M6 socket-head bolt, (8) conical spring washer.}
\end{figure}

Finally we note that this method has also been successfully used to seal an 70~mm diameter AR--coated BK7 optic onto a 70~mm OD pipe on a vacuum chamber for a new cold atom experiment. This seal has been leak tested to less than $10^{-10}$~atm cm$^3$/s. This highlights the versatility of the design which could in principle be scaled up or down to arbitrary sizes for experiments where the use of standard commercial flange sizes is not feasible.

In summary, we have developed a reliably re-useable solder seal
viewport mounted on a modified Conflat half nipple. Problems associated with the
repeated re-use of the viewport have been circumvented with the use
of a decoupled double flange structure. The viewport design has been incorporated into a UHV chamber used for several cold atom experiments.
Furthermore, we have shown that UHV viewports can by created by sealing directly to a vacuum pipe and the use of epoxy on the solder seals gives added protection against small surface imperfections.

\acknowledgements{The authors gratefully acknowledge the assistance
of E.C. Maclagan and S.G. Macleod}


\begin{thebibliography}{99}

\bibitem{muller1988} C. H. Muller III, M. W. Barrett and D. D.
Lowenthal, Rev. Sci. Inst \textbf{59}, 1425 (1988)

\bibitem{nobel1994} A. Nobel and M. Kasevich, Rev. Sci. Inst
\textbf{65}, 3042 (1994)

\bibitem{manuccia1981} T. J. Manuccia, J. R. Peele and C. E.
Geosling, Rev. Sci. Inst \textbf{52}, 1857 (1981)

\bibitem{cox2003} S.G. Cox, P.F. Griffin, C.S. Adams, E. Riis
and D. DeMille, Rev. Sci. Inst \textbf{74}, 3185 (2003)

\bibitem{griff05} P. F. Griffin, K. J. Weatherill and C. S. Adams,
Rev. Sci. Inst. \textbf{76}, 093102 (2005)

\bibitem{griff06} P. F. Griffin, K. J. Weatherill, S. G. MacLeod, R. M. Potvliege and C. S. Adams,
New. J. Phys. \textbf{8}, 11 (2006)

\bibitem{indalloy} Indalloy 165 solder supplied by Indium Corporation of America, 1676 Lincoln Avenue, P.O. Box 269, Utica, NY 13503-0269

\bibitem{belleville} Conical disc springs can be obtained from Belleville Springs Ltd. Arthur Street, Lakeside, Redditch, Worcestershire, B98 8JY, UK and from Key Bellevilles, Inc., 100 Key Lane, Leechburg, PA 15656-9531, USA.

\bibitem{weat08} K. J. Weatherill, J. D. Pritchard, R. P. Abel, M. G. Bason, A. K. Mohapatra and C. S. Adams,
J. Phys. B. \textbf{41}, 201002 (2008)

\bibitem{grimreview} R. Grimm, M. Weiderm{\" u}ller, and Y. B. Ovchinnikov, Adv. Atom. Mol. Opt. Phys \textbf{42}, 95 (2000)

\bibitem{epotech} http://outgassing.nasa.gov/cgi/uncgi/sectionb/sectionb.sh



\end{thebibliography}
\end{document}